\documentstyle[prb,aps,epsf]{revtex}
\newcommand{\be}{\begin{equation}}
\newcommand{\ee}{\end{equation}}
\newcommand{\bea}{\begin{eqnarray}}
\newcommand{\eea}{\end{eqnarray}}

\newcommand{\p}{\partial}
\newcommand{\ri}{{\rm i}}
\newcommand{\re}{{\rm e}}
\newcommand{\rd}{{\rm d}}

\begin {document}
\bibliographystyle {plain}
%\tableofcontents 
%\setlength {\textwidth}{14.5cm
%\setlength {\textwidth}{14.5cm
%\setlength {\textwidth}{14.5cm\pagestyle{nonempty}

\title{ Thermodynamics of  Fateev's models in the Presence of External
Fields
}
\author{Davide Controzzi and  Alexei M. Tsvelik}
\address{Department of Physics, University of Oxford, 1 Keble Road, Oxford, 
OX1 3NP, UK}
\date{\today }
\maketitle

\begin{abstract}
\par
We study the Thermodynamic Bethe
Ansatz equations for a one-parameter quantum field theory recently 
introduced by V.A.Fateev. The presence of chemical potentials produces a kink
condensate that modifies the excitation spectrum. For some combinations
of the chemical potentials an additional gapless mode appears.
Various energy scales
emerge in the problem. An effective field theory, describing
the low energy excitations, is also introduced.
\end{abstract}
\sloppy

%\newpage

\par

\section{Introduction}
Recently V.A.Fateev introduced a series of two dimensional integrable 
deformations of affine Toda theories 
possessing the property of duality\cite{fateev1,fateev2}. These models
exist in two incarnations - fermionic and bosonic - and the strong
coupling limit of one incarnation corresponds to the weak coupling
limit of the other one. Apart from being amusing examples of
integrable field theories, these models may have some interesting
physical applications
\cite{pt1,pt2,ct}.

In this paper we consider in some details one of the simplest models
of that family. The Lagrangian density of the fermionic version of the
theory is given by:
\bea
\label{fermionic}
{\cal L}^{(f)}=\sum_{s =
1,2}\left[\ri\bar\psi_s\gamma_{\mu}\p_{\mu}\psi_s -
\frac{\pi g }{2} (\bar\psi_s\gamma_{\mu}\psi_s)^2\right] 
+ \frac{1}{2}\left[ (\p_\mu\phi_1)^2+(\p_\mu\phi_2)^2 \right] - \\ \nonumber 
M_0\bar\psi_1\psi_1\re^{-\beta\phi_1}-
M_0\bar\psi_2\psi_2\re^{\beta\phi_2} 
- 
\frac{M_0^2}{2\beta^2} \left[ 2 
\re^{\beta(\phi_{2} - \phi_{1})}+
\re^{-
2\beta\phi_1} + \re^{2\beta\phi_2} \right]
\eea 
where $g  = \beta^2/(4\pi + \beta^2)$ and we can identify two types of
fermions $\psi_{1,2}$ and two coupled  Toda chains
 described by the bosonic fields
$\phi_{1,2}$. 

 In the bosonic representation the phonon modes, $\phi_{1,2}$, disappear and   
the  Lagrangian density is :
\be
{\cal L}^{(b)} =\frac{1}{2}\sum_{s = 1}^2\frac{\partial
_{\mu}\chi_s \partial _{\mu}\bar\chi_s }{1+(\gamma /2)^{2}\chi_s
\bar\chi_s }
-
\frac{M_0^{2}}{2}[\chi_1 \bar\chi_1+
\chi_2 \bar\chi_2 +\gamma^2/2(\chi_1\bar\chi_1)(\chi_2 \bar\chi_2)]
 \label{bosonic}
\ee
where the coupling constants of the two models are connected by the
duality relation:
\be
\gamma=4\pi/\beta
\ee 
As we have mentioned above,
the weak coupling limit of the
bosonic theory, $\gamma << 1$ ($1-g << 1$), correspond
to the strong coupling limit of the fermionic one, $ \beta >> 1$. 

The bosonic form of the model is particularly interesting because the
Euclidean version of the theory can be interpreted as Ginzburg-Landau
free energy of two coupled superconductors. We shall discuss this
application in greater details later in the text as well as in a
separate publication.

Models (\ref{fermionic}),(\ref{bosonic}) have  U(1)$\times$U(1) 
symmetry and the corresponding conserved charges
are given by 
\bea
Q_s = \int \rd x\bar\psi_s\gamma_0\psi_s = -\frac{\ri}{2}\int \rd
x\frac{(\bar \chi_s\p_0\chi_s - \chi_s\p_0\bar\chi_s)}{[1 + (\gamma/2)^2
\chi_s\bar\chi_s]}
\eea 
Thus one can introduce two chemical potentials and modify the 
Hamiltonian:
\be
H \rightarrow 
H - h_1Q_1 - h_2Q_2
\ee

The U(1)$\times$U(1) symmetry of the Hamiltonian 
reflects in the symmetry of the particle
multiplets. In absence of chemical potentials the spectrum of the
model consists of fundamental particles carrying quantum numbers of
U(1) groups and their neutral bound states. The two-body scattering
matrix of fundamental particles is given by a tensor product of two 
sine-Gordon 
$S$-matrices multiplied by a CDD-factor responsible for cancellation
of double poles:
\bea
{\cal S}_{12}(\theta) = -\frac{\sinh\theta -
\ri\sin(\pi /\lambda)}{\sinh\theta +
\ri\sin(\pi /\lambda)}
\hat S(\lambda;\theta)\times\hat
S(\lambda;\theta)
\eea
where the $\hat S(\lambda, \theta)$ is the soliton scattering
matrix of the sine-Gordon model with
\be
\lambda=2-g=1+\frac{\gamma^2}{4\pi+\gamma^2}
\label{lambda}
\ee
and is given by
\bea
\hat S_{ab}^{\bar a,\bar b}(\lambda;\theta) =
\re^{\ri\delta_{\lambda}(\theta)}\left 
\{ \delta_{a\bar a}\delta_{b\bar b}\left[
 -
\delta_{ab} + (1 -
\delta_{ab})\frac{\sinh(\lambda\theta)}{\sinh[\lambda(\theta -
\ri\pi)]}\right] \right .\nonumber\\
\left . + \delta_{a\bar b}\delta_{b\bar a}(1 -
\delta_{ab})\frac{\ri\sin(\pi\lambda)}{\sinh[\lambda(\theta -\ri\pi)]}
\right \}
\eea
with
\[
\delta_{\lambda}(\theta) = \int_0^{\infty}
\rd\omega\frac{\sin(\omega\theta)}{\omega}\frac{\sinh[\pi\omega(\lambda^{-1}
- 1)/2]}{\cosh(\pi\omega/2)\sinh(\pi\omega/2\lambda)}
\]

To derive thermodynamic equations we
start from the discrete Bethe Ansatz equations which emerge when the
system is put in a box of size $L$ with periodic boundary conditions:
\bea
\re^{-\ri M\sinh\theta_j L} \xi(a_1,...a_n) & = & 
\prod_{k \neq j} {\cal S} 
(\theta_j - \theta_k)\xi(\bar a_1,...\bar a_n)\label{S}\\
E & = & M\sum_{j = 1}^n\cosh\theta_j
\eea
where $\xi(a_1,...a_n)$ is the wave function of $n$ fundamental
particles in the isotopic space.
The matrix
on the right-hand side of Eq.(\ref{S}) is related  to the  trace of the
monodromy matrix $\tau$:
\be
\prod_{k \neq j}
 {\cal S}(\theta_j - \theta_k) = \mbox{ Tr}
\tau(\lambda = \theta_j; \theta_1,...\theta_n)
\ee
Up to the  scalar factor given by the product of CDD-factors the
 latter matrix
 is equal to a tensor product
of monodromy matrices of the sine-Gordon models:
\bea
{\hat\tau}(\lambda, \theta_1, ... \theta_n) =
{\hat\tau}^{(1)}(\lambda, \theta_1,
... \theta_n)\times{\hat\tau}^{(2)}(\lambda, \theta_1, ... \theta_n)
\eea

The eigenvalues of the sine-Gordon monodromy matrix
  are known, hence
it is straightforward to write down (\ref{S}) in the diagonal form:
\bea
\re^{-\ri M\sinh\theta_j L}  =  \prod_{k \neq j}S_0(\theta_j -
\theta_k) \times  \nonumber \\
\prod_{a = 1}^{m_+}
\frac{\sinh\lambda[(\theta_j - u_a) +
\ri\pi/2]}{\sinh\lambda[(\theta_j
- u_a) - \ri\pi/2]}  \prod_{b = 1}^{m_-}
\frac{\sinh\lambda[(\theta_j - v_b) +
\ri\pi/2]}{\sinh\lambda[(\theta_j - v_b) - \ri\pi/2]}\label{ba1}
\eea
\bea
\prod_{j = 1}^n\frac{\sinh\lambda[(\theta_j - u_a) +
\ri\pi/2]}{\sinh\lambda[(\theta_j - u_a) - \ri\pi/2]} =
\prod_{b = 1}^{m_+}\frac{\sinh\lambda[(u_b - u_a) +
\ri\pi]}{\sinh\lambda[(u_b - u_a) - \ri\pi]}\label{ba2}
\eea
\bea
\prod_{j = 1}^n\frac{\sinh\lambda[(\theta_j - v_a) +
\ri\pi/2]}{\sinh\lambda[(\theta_j - v_a) - \ri\pi/2]} =
\prod_{b = 1}^{m_-
}\frac{\sinh\lambda[(v_b - v_a) +
\ri\pi]}{\sinh\lambda[(v_b - v_a) - \ri\pi]}\label{ba3}
\eea
where:
\bea
S_0(\theta) = \re^{2\ri\delta_{\lambda}(\theta)}\prod_{a = 1}^{n - 1}\frac{\sinh
\theta -
\ri\sin(\pi a/\lambda)}{\sinh\theta +
\ri\sin(\pi a/\lambda)}
\eea

The numbers $m_\pm$ are related to the conserved charges $Q_{1,2}$. As
we shall demonstrate later 
\be
m_\pm=\frac{1}{2} \left[ n-(Q_1\pm Q_2) \right]
\label{pm}
\ee

The simplicity of this formula cannot obscure its remarkable meaning:
propagating particles are superpositions of states localized on
different edges of the stripe. This result is valid for other Fateev's
models where the Toda array consists of more than two elastic chains.

\section{THERMODYNAMIC BETHE ANSATZ (TBA) EQUATIONS}

In order to construct the Thermodynamic Bethe Ansatz (TBA) equations
from (\ref{ba1},\ref{ba2},\ref{ba3})
we
consider the case $1-g=1/\nu$, with integer $\nu\geq 2$, where the
complex solutions of these equations (the
strings) have the simplest classification. For most of the results the
fact that $\nu$ is an  integer is not important and one can generalize
them for arbitrary $g>1/2$ replacing $\nu$ by $(1-g)^{-1}$.

To save space and make our notations more transparent we
shall sometimes  denote the kernels in the integral equations directly
in terms of their Fourier transforms. For example, the convolution of
two functions $g$ and $f$ with the function g having Fourier transform
$g(\omega)$ will be written as:
\be
\int _{-\infty}^{+\infty} d \theta '
g(\theta-\theta')f(\theta')=g*f(\theta)=
\left\{ g(\omega) \right\} *f
\ee
Thermodynamic Bethe Ansatz (TBA) equations have been
introduced  in \cite{fateev1,pt2}. The free energy of the system is
expressed in terms of the excitation energies  $E_n(\theta)$:
\bea
F/L = - T\sum_n M_n\int \frac{d\theta}{2\pi}\cosh\theta\ln[1 + \re^{-
E_n(\theta)/T}]
\eea

 In the specific case
(\ref{fermionic}),(\ref{bosonic})
we have only one bound state with energy $E_1$, and the fundamental
particle with energy $E_0$. The function $E_1$  satisfies the following
equation :
\bea
T\ln[1 + \re^{E_1(\theta)/T}] - 
G_{11}*T\ln[1 + \re^{-E_1(\theta)/T}] = M_1\cosh \theta \nonumber\\
+
G_{10}*T\ln[1 + \re^{-E_0(\theta)/T}],  \label{e1}
\eea
and is directly coupled only to $E_0$, which is determined by the
equation:
\bea
\label{e0}
T \ln (1+\re^{E_0(\theta)/T})-K_{+}*T\ln[1 + \re^{-E_0(\theta)/T}]
= M\cosh\theta -\frac{h_++h_-}{2} \\ \nonumber
+ G_{10}*T\ln[1 + \re^{-E_1(\theta)/T}]
 -   T s*\ln[(1 +
\re^{\epsilon_{-1}(\theta)/T})(1+\re^{\epsilon_{1}(\theta)/T})]
\eea 
The two masses are related by:
\be
M_1=2M \sin[\pi/2(2-g)]
\ee
and the Fourier transform of the kernel is:
\be
K_+ = K_{sym} + 2a_1*s = 1 + \frac{\tanh[\pi\omega/2(\nu +
1)]\sinh[\pi(\nu - 1)\omega/2(\nu + 1)]}{\cosh(\pi\omega/2)}
% =\frac{\cosh(\pi\omega/\lambda)\sinh(\pi(1-1/\lambda)\omega/2)}{
%\cosh(\pi\omega/2)\sinh(\pi \omega/2\lambda)}
\ee
where $K_{sym}$ is related to the two-particle scattering phase :
\be
K_{sym}(\theta)=\delta(\theta)-\frac{1}{2 \pi i}\frac{d}{d\theta} \log
(S_0(\theta)),
\ee
and its Fourier trasform is:
\be
K_{sym}(\omega)=
\frac{\sinh{\pi/2(1-1/\lambda)\omega}
\cosh(\pi\omega/\lambda)}
{\cosh(\pi\omega/2) \sinh(\pi \omega/2\lambda)}.
\ee

The equations for $\epsilon_n$ and $\epsilon_{-n}$ are:
\bea
\epsilon_{-\nu} &  =&  h_{-}\nu/2 -
s*T\ln[1 + \re^{\epsilon_{-( \nu -2)/T}}] \label{tba2}
\eea

\bea
&\epsilon_{-n}&  =  \delta_{n,\nu -1}h_-\nu/2 + s*T\ln[1 +
\re^{-\epsilon_{ n
- 1}/T}][1 + \re^{\epsilon_{- n + 1}/T}]  \label{tba3}\\
& + & \delta_{\nu - 2,n}s* T\ln[1 + \re^{- \epsilon_{-\nu}/T}] + 
%  \varepsilon
\delta_{n,1}s*T\ln[1 + \re^{-E_{0}/T}] \nonumber
\eea
\bea
\epsilon_{\nu} &  =&  h_{+}\nu/2 -
s*T\ln[1 + \re^{\epsilon_{( \nu -2)/T}}] \label{tba4}
\eea

\bea
\label{tba5}
&\epsilon_{n}&  =  \delta_{n,\nu -1}h_+\nu/2 + s*T\ln[1 +
\re^{\epsilon_{n
- 1}/T}][1 + \re^{\epsilon_{n + 1}/T}] 
\\ \nonumber
& + &  \delta_{\nu - 2,n}s* T\ln[1 + \re^{-\epsilon_{\nu}/T}] + 
\delta_{n,1}s*T\ln[1 + \re^{-E_{0}/T}]
\eea
%where $\varepsilon = {\mbox{ sign}}(1/2 - g)$.
The Fourier transforms of the remaining kernels have the form:
%\be
%K(\omega) = \frac{\sinh[\pi\omega(1 -
%g)/2\lambda]\cosh[\pi\omega(\lambda +
%2g)/2\lambda]}{2\cosh\pi\omega/2\cosh(g\pi\omega/2\lambda)
%\sinh(\pi\omega/2\lambda)}
%\label{kernel}
%\ee
\[
s(\omega)=\left[2 \cosh[\pi/2\lambda(1-g)\omega] \right]^{-1}
\]

\[
G_{11}(\omega) =
\frac{4\sinh(g\pi\omega/2\lambda)\sinh[\pi(1 - g)\omega/\lambda]}{
\sinh(\pi\omega/2 \lambda)}
\]
\[
\times\frac{\cosh[(\lambda - 1)\pi\omega/2\lambda]
\sinh[\pi\omega/2\lambda]}{%
\cosh (\pi\omega/2)}
\]
\[
G_{10}(\omega) =
\frac{2\sinh(g\pi\omega/2\lambda)\sinh[\pi(1 - g)\omega/2\lambda]}{
\cosh (\pi\omega/2)}
\]
where $\lambda$ is defined in (\ref{lambda}) and can be also expressed
as $\lambda=1+\nu^{-1}$.

The subsequent analysis will show that the fields $h_\pm$ are linear
combinations of the chemical potentials $h_{1,2}$
\be
h_\pm =h_1\pm h_2
\label{hmp}
\ee
(recall Eq.(\ref{pm})).

\section{PROPERTIES OF THE GROUND STATE}

In this section we consider the zero temperature limit of the TBA
equations. We assume that the fields $h_\pm$ are strong enough to make
$E_0$ negative in some interval $[-B,B]$. Then it is obvious from
Eq.(\ref{e1}) that $E_1>0$:
\be
E_1=M_1\cosh \theta- G_{01}*E_0^{(-)}
\label{gse1}
\ee
From Eqs.(\ref{tba3},\ref{tba5})we deduce that all $\epsilon_{\pm n}$
with $n \neq \nu$ are also positive:
\be
\epsilon_{\pm n}=-a_1*E_0^{(-)}-A_{n,\nu-2}s*\epsilon_{\pm \nu}^{(-)}+
(n-1)h_\pm
\label{neqnu}
\ee
where
\bea
A_{nm}(\omega)&=&2 \coth \Omega \frac{\sinh[(\nu-\max(n,m))\Omega]
\sinh[\min(n,m)\Omega]}{\sinh \nu\Omega} \\ \nonumber 
a_n(\Omega)&=&\frac{\sinh[\pi(\nu-n)\Omega]}{\sinh \nu \Omega}, \;\; 
\Omega=\frac{\pi \omega(1-g)}{2(2-g)} \nonumber
\eea
Substituting Eq.(\ref{neqnu}) into 
Eq.(\ref{tba2},\ref{tba4}) we get the equations
for $\epsilon_{\pm\nu}$:
\be
\epsilon_{\pm\nu}^{(+)}+ \left\{
1-\frac{\sinh[(\nu-2)\Omega]}{\sinh\nu\Omega}
\right\} *\epsilon_{\pm\nu}^{(-)}=h_\pm+\left\{ \frac{\sinh\Omega}
{\sinh \nu\Omega} \right\}* E_0^{(-)}
\label{gs1}
\ee
At last, substituting Eq.(\ref{neqnu}) into Eq.(\ref{e0}) we get:
\be
E_0^{(+)}+K_{sym}*E_0^{(-)}=M\cosh\theta -\frac{1}{2}(h_+ +h_-)+
\left\{ \frac{\sinh\Omega}{\sinh\nu\Omega} \right \}*[\epsilon_\nu^{(-)}
+\epsilon_{-\nu}^{(-)}].
\label{gs2}
\ee
Eqs.(\ref{gs1},\ref{gs2}) determine the structure of the ground
state. We cannot solve them in the general case and therefore we shall
consider only two special cases: a) $h_+<< h_-$; b) $h_+\approx h_-$.

In the first case $\epsilon_{-\nu}>0$ and $\epsilon_\nu$ is negative
inside an interval $[-Q,Q]$, with $Q>>B$ and $Q\rightarrow \infty$ as
$h_+\rightarrow 0$. Since $\epsilon_\nu^{(+)}$ is small it is
convenient to invert the kernel in Eq.(\ref{gs1}) to get:
\be
\left \{ \frac{\sinh\nu\Omega}{2\sinh\Omega \cosh[(\nu-1)\Omega]}
\right \}
*\epsilon_\nu^{(+)}+\epsilon_\nu^{(-)}=\nu h_+/2+\left \{ \frac{1}
{2\cosh[(\nu-1)\Omega]} \right \} *E_0^{(-)}
\label{gs11}
\ee
For $|\theta|>> B$ it is useful to approximate the right hand side of
this equation by its asymptotics:
\be
\left\{ \frac{1}
{2\cosh[(\nu-1)\Omega]} \right\} *E_0^{(-)} \simeq \frac{2\zeta}{\pi}
a_\epsilon
\exp(-\zeta|\theta|)
\label{asimpt}
\ee
where
\be
a_\epsilon=\int_{-B}^B d\theta' E_0(\theta') e^{-\zeta
\theta'}
\label{aeps}
\ee
and
\be
\zeta=\frac{\nu+1}{\nu-1}=\frac{2-g}{g}
\label{zeta11}
\ee

\begin{figure}[here]
\epsfxsize=3.5in
\centerline{\epsfbox{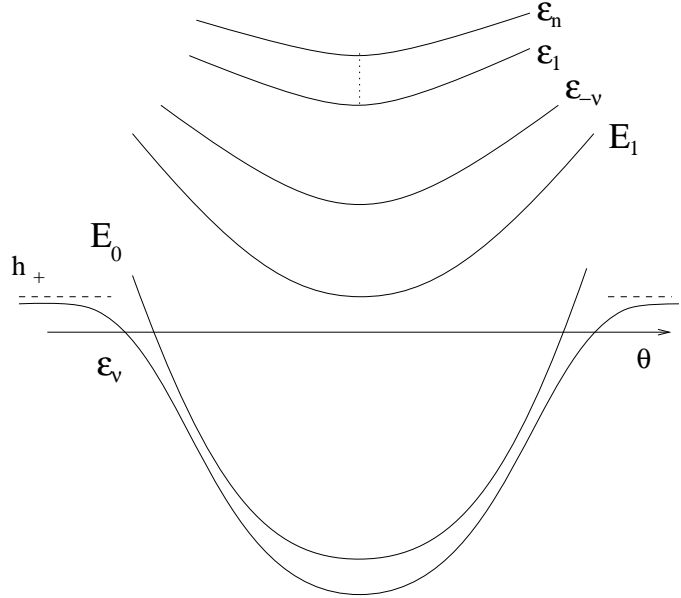}}
\vskip 0.1truein
\protect\caption{Schematic diagram of the ground state energies in
presence of chemical potential for $h_+ <<h_-$ and large B. The same
situation for small B will be presented in section VII
}
\label{en1}
\end{figure}

Analogously one can rewrite Eq.(\ref{gs2}) as:
\be
E_0^{(+)}+K*E_0^{(-)}=M\cosh\theta-h_-/2-\left \{
\frac{1}{2\cosh[(\nu-1)\Omega]} \right \}
*\epsilon_\nu^{(+)}
\ee
where
\be
K(\Omega)=\frac{\sinh\Omega \cosh[(3\nu-1)\Omega]}
{2\sinh\nu\Omega 
\cosh[(\nu-1)\Omega]\cosh[(\nu+1)\Omega] }
\ee
Since $\epsilon_\nu^{(+)}$ is very small, to first approximation this
reduces to:
\be
E_0^{(+)}+K*E_0^{(-)}=M\cosh\theta-h_-/2
\label{gs22}
\ee

For large B we can approximate Eqs.(\ref{gs11}) and (\ref{gs22}) by the
Wiener-Hopf (WH) ones \cite{foz}. The 
solutions for $\varepsilon_0^{(-)}(\theta)=
E_0^{(-)}(\theta+B)$
and $\varepsilon_\nu^{(+)}(\theta)=\epsilon_\nu^{(+)}(\theta+B) $ are:
\bea
\varepsilon_0^{(-)}(\omega)=\frac{h_-}{G_a^{(-)}(\omega)G_a^{(+)}(0)}
\left[ \frac{1}{i\omega+1}-\frac{1}{i\omega+0} \right]
\\
\varepsilon_\nu^{(+)}(\omega)=\frac{1}{G^{(+)}(\omega)} \left[
\frac{h_+}{G^{(-)}(0)}\frac{1}{-i\omega+0}-a_\epsilon
\frac{e^{-\zeta B}}{G^{(-)}(i\zeta)} \frac{1}{\zeta-i\omega} \right]
\eea
where $a_\epsilon$ is defined in Eq.(\ref{aeps}), and 
$G_a^{(+)} (G_a^{(-)})$ and $G^{(+)} (G^{(-)})$ 
are analytic in the upper (lower) half
plane and satisfy the conditions:
\bea
G_a^{(+)}(\omega)G_a^{(-)}(\omega)&=&K(\omega) \\ \nonumber
G^{(+)}(\omega)G^{(-)}(\omega)&=&\frac{\sinh \nu\theta}{2\sinh\Omega
\cosh[(\nu-1)\Omega]}; \nonumber
\eea
$\zeta$ is defined by Eq.(\ref{zeta11}) and 
B is determined by the following relation:
\be
\frac{M e^B}{G^{(+)}(i)}=\frac{h_-}{G^{(-)}(0)}
\ee
To study the low temperature thermodynamic in Section  VI  it is
useful to introduce also the ground state densities for 
$E_0$ and $ \epsilon_\nu$, $\sigma $ and $\Sigma$ respectively,
defined by the following equations:
\bea
\sigma^{(+)}&+&K*\sigma^{(-)}=\frac{M}{2\pi}\cosh \theta
\label{sigma}
\\
\Sigma^{(-)}&+&\left \{ \frac{\sinh \nu\Omega}{2\sinh\Omega
\cosh[(\nu-1)\Omega]} \right \}* \Sigma^{(+)}= \left \{ \frac{1}{2
\cosh[(\nu-1)\Omega]} \right \} * \sigma^{(-)} \equiv \Sigma^0 
\label{Sigma}
\eea
Solutions of the WH equations for $\tilde{ 
\sigma}^{(-)}(\theta)=\sigma^{(-)}(\theta+B)$ and
$\tilde{\Sigma}(\theta)=\Sigma(\theta+B)$ are:
\bea
\tilde{\sigma}^{(-)}(\omega)&=&\frac{h_-}{4 \pi G^{(-)}(\omega)
G^{(+)}(i)(i \omega+1)}\\
\tilde{\Sigma}(\omega)&=&\frac{1}{G^{(+)}(\omega)} \left[
\frac{h_+}{G^{(-)}(0)(-i \omega+0)}-a_\epsilon
\frac{e^{-\zeta B}}{G^{(-)}(i\zeta)(\zeta-i\omega)} \right]
\eea

In the limit b) both $\epsilon_{\pm \nu}$ are positive and we have for
$E_0$:
\be
E_0^{(+)}+K_{sym}*E_0^{(-)}=M\cosh\theta-h
\ee
Again using the WH  method we find the following solution:
\be
\varepsilon_0^{(-)}=\frac{h}{G_b^{(-)}
(\omega)G_b^{(+)}(0)(i\omega+1)(i\omega+\delta)}
\ee
with $G_b^{(+)}(\omega)G_b^{(-)}(\omega)=K_{sym}(\omega)$.
Substituting this solution in Eq.(\ref{gs1}) we obtain the following
estimate for the gap in $\epsilon_{\pm \nu}$:
\be
M_\nu \sim h(M/h)^{(2-g)/2}
\label{gap}
\ee
In the limit $\nu^{-1}\rightarrow 0 ( g\rightarrow 1)M_\nu \sim \sqrt{Mh}$.

A schematic diagram of these results is presented in Fig.1 and Fig.2
for the case a) and b) respectively.

\begin{figure}[here]
\epsfxsize=3.5in
\centerline{\epsfbox{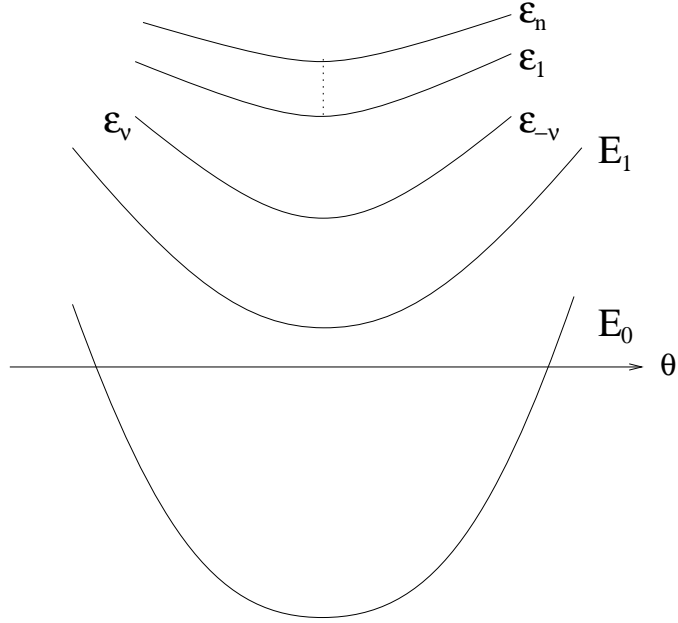}}
\vskip 0.1truein
\protect\caption{Schematic diagram of the ground state energies in
presence of chemical potential for $h_+ \approx  h_- $ and large B. 
}
\label{en2}
\end{figure}

\section{THE LIMIT OF FREE BOSONS}
 
To check the validity of TBA equations (\ref{e1}-\ref{tba5}) 
and establish the connection
between the fields $h_\pm$ and the chemical potentials $h_{1,2}$ we
consider the free boson limit of the TBAs.
It is 
realized in the case $g \rightarrow1$ or $\nu^{-1} \rightarrow 0 $.
In this case the kernels in Eqs.(\ref{tba2}-\ref{tba5}) become
proportional to the delta functions
\[
s(\theta) \rightarrow \frac{1}{2}\delta(\theta)
\]
and $K_+ \rightarrow I, G_{11} = G_{01} \rightarrow 0$, such that all
TBA equations become algebraic.
The mass of the bound state in this limit is $M_1=2M$ and the Free Energy
is equal to:
\be
F/L=-\frac{TM}{2\pi} \int d\theta \cosh\theta \ln \left \{ \left[
1+e^{-E_1/T} \right]^2 \left[ 1+e^{-E_0/T} \right] \right \}
\ee

 The solution of TBA equations 
is expressed in terms of $\xi = M\cosh\theta/T$:
\be
1 + \re^{- E_1/T} = (1 - \re^{-2\xi})^{-1}\label{E1}
\ee
\bea
1 + \re^{\epsilon_{\pm n}/T} = \left[\frac{\sinh (h_{\pm}n +
a_{\pm})/2T}{\sinh h_{\pm}/2T}\right]^2
\eea
where
\bea
\frac{\sinh a_{\pm}/2T}{\sinh h_{\pm}/2T} = \sqrt{1 + \re^{- E_0/T}}
\eea
Substituting this into Eq.(\ref{e0}) we get
\[
\left\{\cosh(h_+/2T)\sqrt{1 + \re^{- E_0/T}} + [\cosh^2(h_+/2T) +
\sinh^2(h_+/2T)\re^{- E_0/T}]^{1/2}\right\}\times
\]
\bea
\left\{\cosh(h_-/2T)\sqrt{1 + \re^{- E_0/T}} + [\cosh^2(h_-/2T) +
\sinh^2(h_-/2T)\re^{- E_0/T}]^{1/2}\right\} = \re^{\xi - E_0/T}
\eea

which has the following solution:
\bea
1 + \re^{- E_0/T} = \frac{\sinh^2\xi}{(\cosh\xi -
\cosh(h_+/2T)\cosh(h_-/2T))^2 - \sinh(h_+/2T)\sinh(h_-/2T)}
\eea
Combining it with (\ref{E1}) in the Free Energy, we get
\be
F/L=-\frac{T}{2\pi} \int d \theta M \cosh \theta  \ln({\cal F})
\ee
\bea 
{\cal F} =  (1 - \re^{-2\xi})^{-2}\frac{\sinh^2\xi}{(\cosh\xi -
\cosh(h_+/2T)\cosh(h_-/2T))^2 - \sinh^2(h_+/2T)\sinh^2(h_-/2T)}
\eea
\[
=
\left \{ [1-e^{-\xi-(h_++h_-)/2T}][1-e^{-\xi+(h_++h_-)/2T}]
[1-e^{-\xi-(h_+-h_-)/2T}][1-e^{-\xi+(h_+-h_-)/2T}] \right \}^{-1}
\]
The free boson limit of TBAs (\ref{e1}-\ref{tba5}) does reproduce the 
free energy of two non interacting complex bosonic fields. This
indicates that the TBA equations are correct. Another important result
is the confirmation of Eqs.(\ref{pm},\ref{hmp}).

\section{SEMICLASSICAL ANALYSIS OF THE LIMIT OF SMALL $\gamma$}

Let us consider the Euclidean form of the 
Lagrangian (\ref{bosonic}).
Rescaling the fields,  $\chi_s \rightarrow (\gamma/2) \Delta_s $, it becomes:
\bea
{\cal L}^{(b)}=\frac{2}{\gamma^2} \left [ \sum_{s =
1}^2\frac{\partial
_{\mu}\Delta_s \partial _{\mu}\bar\Delta_s }{1+ \Delta_s
\bar\Delta_s }
+
M^{2}\left \{ \Delta_1 \bar\Delta_1+
\Delta_2 \bar\Delta_2 +2(\Delta_1\bar\Delta_1)(\Delta_2
\bar\Delta_2)\right \}\right]
%\equiv E(\Delta_s,h_s)
\label{classical}
\eea
Then in
the limit $\gamma \rightarrow 0 $ the quantum fluctuations are
suppressed and one can approximate the ground state energy as minimum of
the functional 
\be
E(\Delta_s,h_s)=\frac{2}{\gamma^2} \left [  M^{2} \left 
\{\Delta_1 \bar\Delta_1+
\Delta_2 \bar\Delta_2 +2(\Delta_1\bar\Delta_1)(\Delta_2
\bar\Delta_2) \right\} - \sum_{s=1,2} \frac{h_s \Delta_s \bar\Delta_s}{
1+ \Delta_s
\bar\Delta_s } \right ]. 
\ee
We will not address this problem,
already studied by Fateev \cite{fateev1}, but we will use this
approximation to
to  get more insight to the problem and give an intuitive
picture of the low energy physics.
Let us  perform the
following transformation in the form (\ref{classical}):
\be
\Delta_s= \sinh\rho_s e^{i\varphi_s}
\ee
Under this  transformation  the measure transform
in $D\rho D\varphi \sinh(2\varphi)$
and the Lagrangian density in the new fields in presence of chemical
potentials assume the form:
\bea
{\cal L}^{(b)}= \frac{2}{\gamma^2}
\{ \sum_{s=1,2}[ (\p_\mu \rho_s)^2+
\tanh ^2 \rho_s(\p_\mu \varphi_s)^2- h_s \tanh ^2\rho_s \p_0 \varphi_s]
\\ \nonumber
+M^2  [\sinh ^2 \rho_1+\sinh^2  \rho_2
+2 \sinh^2\rho_1
\sinh ^2 \rho_2] \}
\eea
For $h_1=h$ and $h_2=0$, which corresponds to the situation b in
section III, we get:
\bea
{\cal L}^{(b)}=\frac{2}{\gamma^2} \{ (\p_\mu \rho_1)^2+
\tanh ^2 \rho_1 [(\p_\mu \varphi_1)^2- h\p_0 \varphi_1]+
 (\p_\mu \rho_2)^2+ \\ \nonumber
\tanh ^2 \rho_2 (\p_\mu \varphi_2 )^2
+ M^2  (\sinh ^2 \rho_1+\sinh^2  \rho_2+
2 \sinh^2\rho_1
\sinh ^2 \rho_2) \}
\eea
The term $h(\p_0 \varphi_1)$ can be absorbed by the shift:
\be
\varphi_1 \rightarrow \varphi_1-(h/2)t
\ee
which generate also an additional term $ - h^2/4$.
Then the Lagrangian density becomes:
\be
{\cal L}^{(b)}=\frac{2}{\gamma^2} \{ 
(\p_\mu \varphi_1)^2 +(\p_\mu
\varphi_2)^2+\tanh^2\rho_1(\p_\mu \varphi_1)^2+\tanh^2 
\rho_2(\p_\mu \varphi_2)^2 
+
V^b_{eff}(\rho_1,\rho_2) \}
\ee
where:
\be
V^b_{eff}(\rho_1,\rho_2)=V_1(\rho_1)+M ^2 \sinh ^2 \rho_2+2 M^2 \sinh^2\rho_1
\sinh^2 \rho_2
\ee
and
\be
V_1(\rho_1)=M^2 \sinh^2 \rho_1-h^2/4\tanh^2\rho_1
\ee
For $h/2 > M$, $ V^b_{eff}(\rho_1,\rho_2)$ 
has a minimum at $\rho_1=\rho_0$ and $\rho_2=0$ where 
$\exp 2\rho_0 =\sqrt{8}(h/M)$. In the vicinity of the minimum we have:
\be
V_{eff}(\rho_1=\rho_0+x,\rho_2=y)\approx hM(\sqrt{2}y^2+\sqrt{8}x^2)
\ee
which gives the masses going like  $\sqrt{hM}$, 
in accordance with (\ref{gap}).

Thus when the  chemical potential exceeds the threshold 
the kinks condense and  gapless Goldstone  mode appear. To see
this explicitly  we write $
\rho_1=\rho_0+x$ 
and keeping only quadratic terms in $x$ obtain the following
expression for 
the Lagrangian:
\bea
{\cal L}^{(b)}\sim \frac{2}{\gamma^2} \{(\p_\mu x)^2
+(Mh/2) x^2+\tanh^2
\rho_0 (\p_\mu \varphi_1)^2+ \\ \nonumber
(\p_\mu \rho_2)^2+\tanh^2\rho_2(\p_\mu
\varphi_2)^2+\sinh^2\rho_2 h/M \}
\eea
Here  we can identify  a  gapless mode  $\varphi_1$ with velocity
equal to the bare one, a gapful field $x$ with the mass $m_x^2=hM/2$
and an effective integrable field theory
described by the Lagrangian density:
\be
{\cal L}_2=\frac{1}{2}(\frac{2}{\gamma})^2 \{ M \sinh^2 \rho_2 + (\p_\mu
\rho_2)^2 +\tanh^2\rho_2 (\p_\mu \varphi_2)^2 \}
\ee

In the case $h_1\sim h_2$  corresponding  to the situation a) of
Section III,  the effective potential acquires  the form:
\be
V^a_{eff}(\rho_1,\rho_2)=V_1(\rho_1)+V_1(\rho_2)+2 M^2 \sinh ^2 \rho_1 
\sinh^2\rho_2
\ee
Repeating the same analysis one finds two independent gapless modes
with the same velocity. Then the semiclassical approximation gives a
correct qualitative description of the low energy behavior of the system. As
a byproduct we obtain another confirmation of (\ref{hmp}).

\section{LOW ENERGY PHYSICS: MASSLESS MODES AND ENERGY SCALES}

Let us go back to the zero temperature limit described in Section
III. As we have shown, in the case $h_1 \approx h_2$ there are two
soft modes: $E_0$ and $\epsilon_\nu$. At $|\theta|< B$ the mode
$\epsilon_\nu$ closely follows $E_0$; at$|\theta| >> B$ its behavior
is determined by the asymptotics (\ref{asimpt}) and this mode becomes
independent. The temperature scale below which the two modes decouple
can be determined by the value of of $\epsilon_\nu$ at $\theta \sim
B$:
\be
T_{sep} \sim \epsilon_\nu^{(-)}(B)
\label{tsep}
\ee
For low temperatures, $T<T_{sep}$,  the two modes  give independent
contribution to the thermodynamics. To calculate the low temperature
free energy:
\be
F/L=-\frac{T}{2\pi} M \int d\theta \cosh\theta
\ln[1 + \re^{-E_0(\theta)/T}]
\label{free}
\ee
we use the Eqs. (\ref{gs11},\ref{gs22}) which at finite temperature
have the form:
\bea
T \ln (1+e^{E_0/T})-K*T \ln (1+e^{-E_0/T})=M \cosh
\theta-\frac{h_-}{2}-\left \{ \frac{1}{2 \cosh(\nu-1)\Omega} \right
\}*T \ln(1+e^{\epsilon_\nu/T}) 
\label{te0} \\
\left \{ \frac{\sinh\nu\Omega}{2 \sinh\Omega \cosh(1-\nu)\Omega} \right \}*
T \ln (1+e^{\epsilon_\nu/T})-T \ln (1+e^{-\epsilon_\nu/T})
=\frac{\nu h_+}{2}-a_\epsilon
\label{tenu}
\eea
where $a_\epsilon$ is defined in Eq.(\ref{asimpt}).
To isolate the contributions that vanish at T=0 we rewrite (\ref{te0})
like:
\be
E_0^{(+)}+K*E_0^{(-)}+ (I-K)* T \ln(1+e^{-|E_0|/T})=\mbox{r.h.s. of Eq.(\ref{te0})}
\ee
where $I$ is the identity operator. Using Eq.(\ref{sigma}) we can rewrite
the free energy as:
\be
F/L=-\frac{T}{2\pi} M \int d\theta \cosh\theta (E_0^{(-)}+ T
\ln(1+e^{-|E_0|/T}))\equiv f_0+f_\nu
\ee
where 
\bea
f_0=-T \int d\theta \sigma (\theta) 
\ln(1+e^{-|E_0|/T})\approx -\pi^2 T^2/3V_c
\\
f_\nu=-T \int d\theta \left \{ \frac{1}{2\cosh (\nu-1)\Omega} \right
\}
\sigma^{(-)} (\theta)  T
\ln(1+e^{\epsilon_\nu/T}) \approx -\pi^2 T^2/3V_s
\eea
The velocities of the two modes are determined by :
\bea
V_c = \sum_{s = \pm}|\frac{\p E_0(\theta)}{\p\theta}|
[2\pi\sigma^{(s)}(B)]^{-1}_{\theta = sB} \\
V_s = \sum_{s = \pm}|\frac{\p\epsilon_\nu(\theta)}{\p\theta}|
[2\pi\Sigma^{(s)}(B)]^{-1}_{\theta = sB}
\eea
For large B all the quantities appearing in these  equations can be
calculated using  the Wiener-Hopf  solutions obtained in Section III via the
following relations:
\bea
\left |  d{E_0}^{(\pm)}/ d \theta  \right |_{\theta=B}=
\left |  d{\varepsilon_0}^{(\pm)}/ d \theta  \right |_{\theta=0} & = &
\lim_{\omega \rightarrow \infty} \pm (i \omega)^2 {\epsilon}^{(\pm)}
(\omega) \nonumber \\
\sigma^{(\pm)}(B)=
 {\tilde \sigma}^{(\pm)}(0) & = & \lim_{\omega \rightarrow \infty} \pm i
 \omega {\tilde \sigma}^{(+)} (\omega) \nonumber
\eea
and analogous relations for $\epsilon_\nu$ and $\Sigma$.
The results are:
\be
V_c=V_s=1
\ee

The fact that the two velocities are the same 
seems to be peculiar to the model in consideration and not
a general property of the family of Fateev's models. In general
$V_s$ depends on g and on the number of Toda chains.

Thus we have shown that the presence of chemical potentials introduces
additional energy scales  in the problem with
non trivial crossovers.
At temperatures  $T\ll h$, 
 $\epsilon_{-n}$ and $\epsilon_{n} \;(n \neq\nu)$ 
modes decouple from the other  modes. The
same happens for $E_1$ at energies smaller then their gap and in
some physical regions there is also an  intermediate energy scale, as
we will see in the next Section. As
one 
lowers the temperature below  $T_{sep}$ the gapless modes decouple from
each other and the picture of two  non-interacting Luttinger liquids
emerges.

\section{WEAK COUPLING LIMIT: NON PERTURBATIVE RESULTS}

Now we wish  to focus our  attention on the limit
$g \rightarrow 1$ which corresponds to the weak-coupling limit of the
bosonic theory.
Naively one can imagine to obtain perturbative results. It turns out 
however that, in the presence of the chemical potential, 
 this is not the case, the reason being  that  the kinks condensate survives
and affects the excitation spectrum. A similar effect is present in the
Klein-Gordon limit of the sine-Gordon model \cite{js}.
 
To be specific let us consider the situation a) of Section III, all the
energies $\epsilon_n \; (n\neq \nu) $ 
have gaps of the order of $h_-$ and can be omitted at temperatures 
$T<<h_-$. 
To obtain information on the excitation spectrum one has to study the
fundamental equations (\ref{gse1}) and (\ref{gs22}). 
The standard
relativistic dispersion law: $E_1(\theta)=M_1 \cosh \theta$, can be
modified by the last term in (\ref{gse1}) which emerge in presence of
kinks.

In the  limit $g\rightarrow 1$,  
the kernels simplify $K(\omega)=K_0(\omega)(1+O((1-g)^{2}))$, and
$G_{10}(\omega)=G_0(\omega)(1+O((1-g)^{2}))$, where 
\bea
\label{k00}
K_0(\omega)=\frac{4\pi^2}{\beta^2\lambda}\omega \frac{\cosh(\pi
\omega(\lambda+2)/2\lambda)}{\cosh(\pi\omega/2)\sinh(\pi\omega/\lambda)}\\
\nonumber
=\frac{\pi}{\lambda}(1-g) \omega \frac{\cosh(\pi
\omega(\lambda+2)/2\lambda)}{\cosh(\pi\omega/2)\sinh(\pi\omega/\lambda)}
\eea
and
\be
G_0=(1-g)\frac{\pi\omega}{\lambda}\frac{
\sinh(\pi\omega/2\lambda)}{\cosh(\pi\omega/2)}
\ee
Again we can study TBA equations with the kernel $K_0(\omega)$ and
$G_0(\omega)$ in the
limits of large and small B.
We focus the attention on this second case   where
the effects we want to consider can be seen more clearly. 
To study Eq.(\ref{gse1}) for small $B$  we need the kernel $K_0(\theta)$
only for  small $\theta$. This is equivalent to
approximating Eq.(\ref{k00}) for large $\omega$, as:
\be 
K_0(\omega)\simeq \frac{2\pi}{\lambda}(1-g)|\omega|
\label{k0}
\ee
Approximating also the rhs of Eq. (\ref{gs22})
for small $\theta$ we can rewrite it as:
\be
K_0\tilde *E_0=M\theta^2/2-H
\ee 
where 
$H=h_- /2 -M$ and
$\tilde *$ denote the convolution of support $(-B,B)$ and $K_0$ is
the Fourier transform of (\ref{k0}). 
One can absorb the constant part of the kernel in $E_0$ introducing
the new function 
$\tilde E_0=\frac{2\pi}{\lambda}(1-g) E_0$. Equation (\ref{gs22}) then 
becomes:
\be
\tilde K_0 \tilde * \tilde E_0=M \theta^2/2-H, 
\label{wc1}
\ee
$\tilde K_0$ is the singular integral operator
corresponding to the kernel $\tilde{K}_0(\omega)=|\omega|$ and defined:
\be
\tilde K_0*\tilde E_0(\theta)=-\frac{1}{\pi} P \int_{-B}^{B}
\frac{1}{(\theta-\theta')^2}\tilde E_0(\theta')
d\theta'=\frac{1}{\pi}\frac{\partial}{\p\theta}
P\int_{-B}^{B} 
\frac{1}{(\theta-\theta')}\tilde E_0 (\theta')  \rd \theta '
\ee

\begin{figure}[here]
\epsfxsize=5in
\centerline{\epsfbox{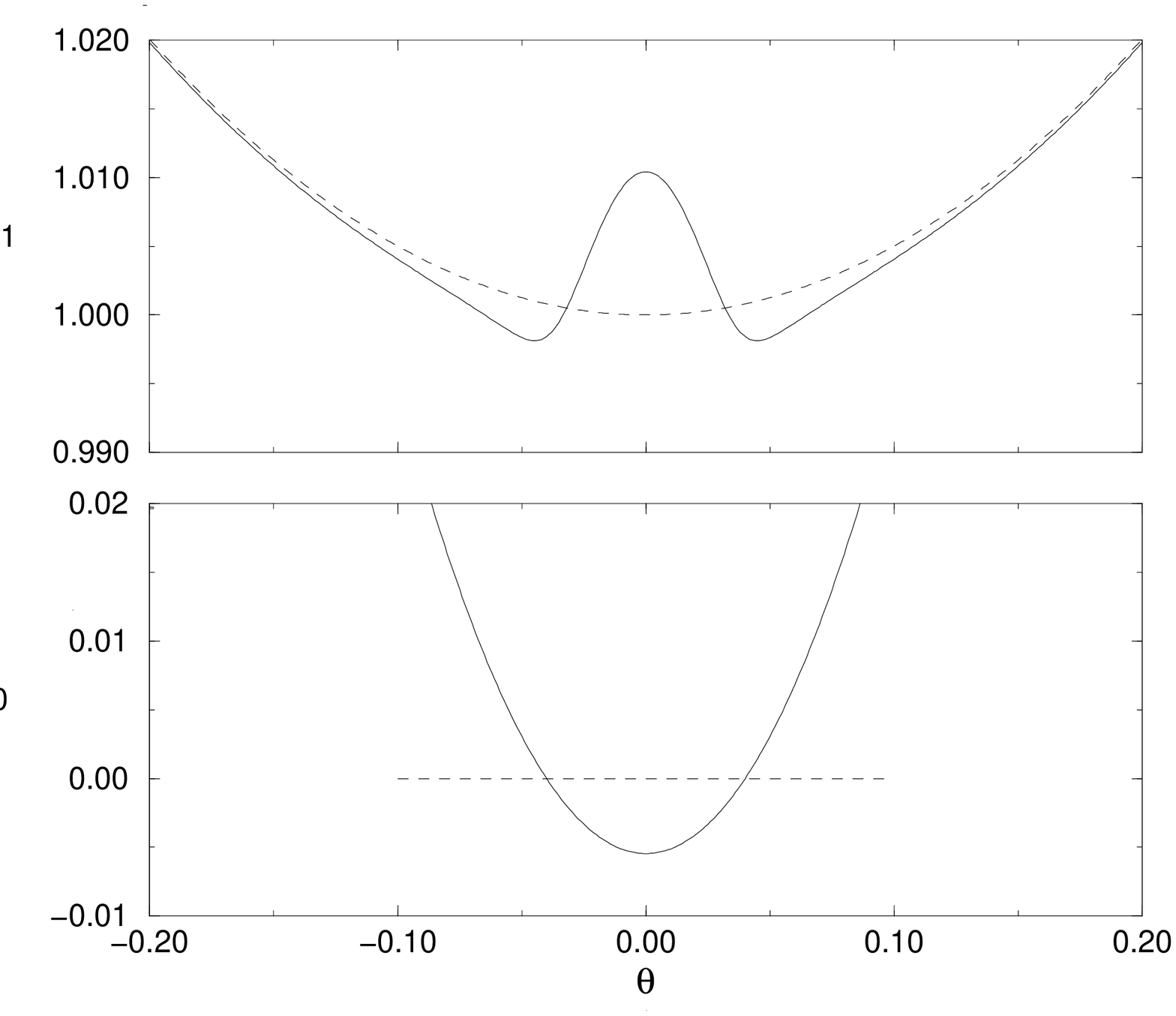}}
\vskip 0.1truein
\protect\caption{Dispersion relation $E_1$ (upper figure)
in the presence of kinks
for $B << 1$ and  $g =0.02$,
from the numerical solution of Eq.(\ref{gse1}). The dotted line
represents
the solution in the absence of kink condensate. In the lower figure
we plot  the gapless mode $E_0$.
}
\label{disp}
\end{figure}

Using this  form of the kernel, Eq. (\ref{wc1}) 
can be reduced to the canonical form:
\be
P\int_{-B}^{B} \rd \theta ' \frac{1}{(\theta-\theta')} V(\theta')=g(\theta)
\ee
where we have introduced:
$ V(\theta)=\p \tilde E_0(\theta)/\p\theta$ 
and $g(\theta)=\pi (M \theta^2/2-H)$.
This equation has the following general solution\cite{eq}:
\be
V(\theta) = \frac{1}{\pi^2}\sqrt{B^2 - \theta^2}
P\int_{-B}^B \rd\theta'\frac{g(\theta')}{\sqrt{B^2 -
\theta'^2}(\theta' - \theta)}
\ee
which gives:
\be
E_0(\theta) = - \frac{M\lambda}{3\pi^2(1 - g)}(B^2-\theta^2)^{3/2}
\ee
where $B^2/4=\frac{h_-/2-M}{M} \equiv \Delta$.
Analogously the ground state density (\ref{sigma}) turns out to be:
\be
\sigma(\theta) = \frac{M\lambda}{2\pi^3(1 - g)}\sqrt{B^2-
\theta^2}
\ee
From here we find the  relationship between $h_-$ and the charge:
\be
Q=\frac{h_-/2-M}{2\pi(1-g)}
\ee
The Fermi energy, $E_F=E_0(\theta)$ as function of the charge if
therefore:
\be
E_F=\frac{8M}{3\pi^2}(1-g)^{1/2} \left [ \frac{2\pi}{M}Q \right ]^{3/2}
\ee

We can now  use the above results to  solve Eq(\ref{gse1}). Rewriting
the Fourier transform of the kernel $G_0$ as: 
\be
G_0\sim \frac{2}{\pi}\frac{\p}{\p\theta}\frac{(\pi/2(1-g))
\sinh\theta}{
\sinh^2\theta+(\pi/2(1-g))^2 \cosh^2\theta}
\ee
it is possible to extract the asymptotic behavior  of $E_1$:
\be
E_1=\left\{ \begin{array}{ll}
M_1\cosh\theta+M/3 \cos(2\sin^{-1}(\theta/B)) , \theta \ll B \\
M_1\cosh\theta+A e^{-\theta},  \theta \gg B
\end{array} \right.
\ee
The numerical solution is shown  on Fig. 3.
One can clearly see that even in the weak coupling limit the 
dispersion law is drastically modified at small $\theta$.

 This   analysis   is supposed to clarify   the subject  
of  different  energy scales present in the model (cfr.Fig.4). 
At very small doping,
$Q \ll Q_1 \equiv  M(g-1)^{-1/3}$, $E_F$ is smaller then M and there
are two scales besides $h$:  M
and $E_F$ itself. The mode $E_1$ is completely decoupled at low
temperature.  As you increase the doping you reach an intermediate
region, characterized by
$M(1-g)^{-1/3} \ll Q \ll Q_2 \equiv  M(1-g)^{-1}$, 
where $E_F \simeq M$ but still much smaller then $E_1$. 
At very large doping we find a very interesting region. As we have
seen in Sec. III at large Q (or equivalently at large B) 
$E_F \simeq  Q$, while $E_1(0)
\simeq \sqrt{Q}$. Then there is a regime in which $E_F$  is
bigger then the gap of the $E_1$-mode, and  there will be  temperatures, $T
\ll E_F$  for
which the  $E_1$-mode becomes soft. 

\begin{figure}[here]
\epsfxsize=4in
\centerline{\epsfbox{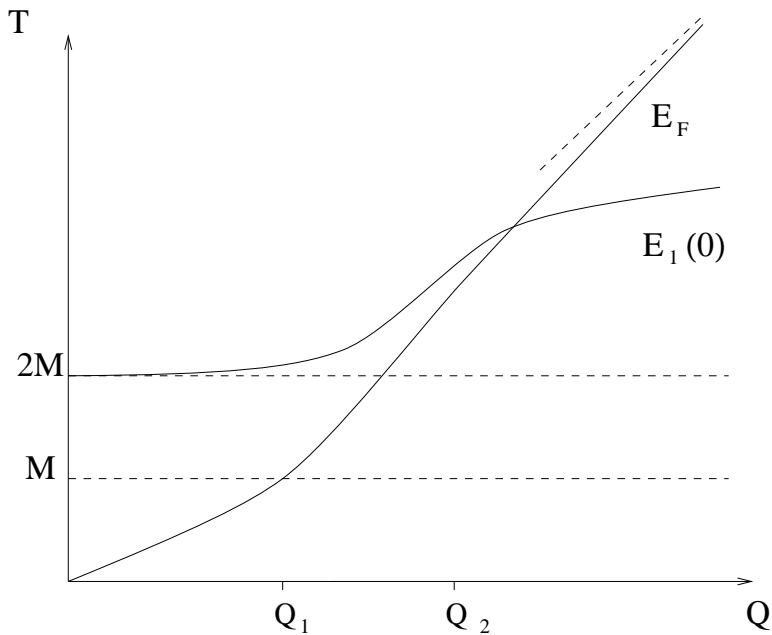}}
\vskip 0.1truein
\protect\caption{A Schematic diagram describing  different
energy scales as a function of the charge. We can recognize three
distinct regions as described in Sec. VII.
}
\label{disp}
\end{figure}

\section{DISCUSSION}

We have considered in some detail the thermodynamics of one of the
integrable field theories recently introduced by Fateev. Most of the
results are quite general and remain valid also for other models of
the family. These
models are interesting for various reasons. First of all they have
peculiar mathematical properties mostly related to a dual
representation of the theory.
In addition they probably can find application to various physical
systems. Considering the model as a (1+1)-dimensional Quantum Field
Theory
one can interpret it  as two one
dimensional conductors coupled via phonon interaction. 
The problem of quasi one dimensional systems with electron-phonon
interaction is a very important one and cannot be approached with
perturbative methods, it is then very important to have exact results
for some specific model.
This
interpretation of the model and the possible application to physical systems 
have been discussed in ref. \cite{pt2}, 
for physical regimes different from the one considered here (i.e. for
$g<1/2$). 
On the other
side, the Euclidean version of the model, in the bosonic
representation,
describes an effective
Landau-Ginzburg theory of two coupled superconductors. In the specific
model considered in this work, the coupling between the 
superconductors is quite simple. Nevertheless  for other models of the Fateev's
family, where coupling between the fermionic modes is achieved via 
a higher number of Toda chains, the resulting coupling between the two
superconductors is via elastic modes and then can be
interpreted as   
two layered superconductors separated by  an insulating stratum. 
In relation to the high-T$_c$ (cuprate) superconductors an interesting
problem is to study the dependence of the superconducting properties
on the number of insulating layers between the superconducting planes. We
will address this problem in a separate publication.

Let us give a brief summary of our results.  As shown in Sec.VII the
results are non-perturbative also in the weak coupling regime  and
could only be obtained through exact or non-perturbative
methods. This is related to the fact that the  chemical
potentials generate a  kink condensate that survives in the weak coupling
limit.
As we have shown, for some combinations of the chemical
potentials two  gapless modes emerge and various energy scales
appears, which characterizes crossovers between different regimes.
Probably the most interesting region of the phase diagram
is obtained at large doping where
the $E_1$-mode  become soft at temperatures for which the $E_0$-one is
still frozen. It deserves further investigation.
Another important result is the one  
contained in Eq.(17). This equation tells us that charged particles
in the Fateev's model are linear combinations of particles located on
the edges, which is important for qualitative 
understanding of the physics of the
model.

%\newpage

\end{document}